\documentclass[aps,prl,twocolumn,groupedaddress]{revtex4-1}
\usepackage[rgb,dvipsnames]{xcolor}
\usepackage{latexsym}
\usepackage{amsmath, amsthm, amssymb}
\usepackage{epsfig}
\usepackage{latexsym}
\usepackage{bm}

\newcommand{\eq}[1]{Eq.~(\ref{#1})}
\newcommand{\fig}[1]{Fig.~\ref{#1}}

\newcommand{\CMT}[1]{\textcolor{blue}{{\it #1}}}     
\newcommand{\TT}[1]{{\ttfamily #1}}

\DeclareFontFamily{OT1}{pzc}{}
\DeclareFontShape{OT1}{pzc}{m}{it}{<-> s * [1.10] pzcmi7t}{}
\DeclareMathAlphabet{\mathpzc}{OT1}{pzc}{m}{it}

\begin{document}

\title{Milling and meandering: Flocking dynamics of stochastically interacting agents with a field of view}
\author{Trilochan Bagarti and Shakti N. Menon}
\affiliation{The Institute of Mathematical Sciences, CIT Campus, Taramani, Chennai 600113, India}

\date{\today}

\begin{abstract}
We introduce a stochastic agent-based model for the flocking dynamics of
self-propelled particles that exhibit velocity-alignment interactions with
neighbours within their field of view. The stochasticity in the dynamics 
of the model arises purely from the uncertainties at the level of interactions. 
Despite the absence of attractive forces, this model gives rise to a wide array
of emergent patterns that exhibit long-time spatial cohesion. In order to gain
further insights into the dynamical nature of the resulting patterns, we
investigate the system behaviour using an algorithm that identifies
spatially distinct clusters of the flock
and computes their corresponding angular momenta. Our results
suggest that the choice of field of view is crucial in determining the resulting
emergent dynamics of stochastically interacting particles.	
\end{abstract}

\maketitle

The collective movement of large groups of microorganisms, insects, birds,
and mammals are amongst the most spectacular examples of self-organized 
phenomena in the natural world~\cite{Vicsek2012, Sumpter2006}. Species across
a range of length scales exhibit a rich variety of collective patterns of motion
that are united by similar underlying characteristics~\cite{Parrish1997,Menon2010}.
Advances in experimental techniques for investigating flocking~\cite{Cavagna2018}
has sustained interest in uncovering the principles that underpin this emergent
phenomenon. For instance, recent experiments have demonstrated that pairwise
interactions motivated by biological goals play a crucial role in determining
insect swarming patterns~\cite{Puckett2015}. Flocks may fundamentally be
viewed as dry active matter, namely systems of self-propelled particles that do
not exhibit conservation of momentum~\cite{Marchetti2013}, and their dynamics can
be understood as a process similar to the long-range ordering of interacting
particles~\cite{Cavagna2014}. Following the seminal work of
Vicsek \emph{et al.}~\cite{Vicsek1995,Ginelli2016}, the dominant paradigm in
models of flocking is that stochasticity in the dynamics can be accounted for
through external noise (either additive or multiplicative). However, this
approach is only truly appropriate for situations such as a system of Brownian
particles, where fluctuations arise from the surrounding media. In contrast,
experimental evidence suggests that the dominant contribution to the
stochasticity in flocks arises from variability in the behaviour of individual
particles~\cite{Aplin2014,Delgado2018}. Furthermore, the collective dynamics of
a swarm is known to be density-dependent~\cite{Buhl2006,Yates2009}, which
tacitly suggests that variations in individual behaviour may have a cumulative
impact. Indeed, flocks may exhibit ordered macroscopic dynamics even if the
behaviour of individual particles is subject to noise~\cite{Niizato2018}.
Hence, it is of significant interest to consider the emergent flocking
behaviour in a system where stochasticity arises purely from the
uncertainties at the level of inter-particle interactions.

In situations where individual particles are unable to uniformly survey their 
neighbourhood due to physiological or other constraints, their interactions would
be limited to neighbours that lie within a field of view~\cite{Hemelrijk2012}.
It has been observed that even a minimal assumption of fore-aft asymmetry
can significantly impact the collective dynamics of a flock~\cite{Chen2017}.
Furthermore, a range of flocking patterns can be observed in a system with
position-dependent short range interactions restricted by a vision cone
~\cite{Barberis2016}.
Recently, we demonstrated that similar constraints on the field of view
of a particle in a two-dimensional lattice model of flocking
can yield a jamming transition even at extremely low particle densities~\cite{Menon2017}.
However, the role of a field
of view on the dynamics of particles that undergo stochastic velocity alignments
remains an open question.
Moreover, while certain types of position-dependent interactions can facilitate
cohesion in a flock~\cite{Gregoire2003,Gregoire2004}, it is intriguing to consider
how this outcome might be achieved with velocity alignments alone.
Furthermore, while some flocking models have  incorporated the
acceleration of particles to describe short-term memory~\cite{Szabo2009},
collision avoidance~\cite{Peng2009}, consensus decision
making~\cite{Bhattacharya2010} and other experimentally observed
features~\cite{Mishra2012}, the role of position-independent stochastic
acceleration remains to be established.

In order to address these questions, we propose in this article a novel
paradigm for flocking in which long-time spatial cohesion can emerge through a
stochastic acceleration, despite the absence of attractive forces or explicit
confinement.
While there have been previous attempts at incorporating stochasticity arising
from an individual particle's evaluation of their interactions with agents
in their neighbourhood (for example \cite{Chate2008}), here we explicitly
consider a situation where, at each instant, particles interact with a single
randomly chosen neighbour in their field of view.
We assume that the interaction between a chosen pair of particles
depends only on their respective velocities, in contrast to the typical assumption
of two-body or mean-field interactions that depend on the relative positions
of particles. Furthermore, while most previous flocking models account for stochasticity
through an external noise, here it is a consequence of uncertainty in velocity
alignments. This leads to a variety of emergent collective dynamical patterns
whose spatio-temporal characteristics vary significantly. Finally, in order to
classify these patterns in a unified manner, we present a cluster-finding
algorithm that determines the spatially distinct clusters of the flock and
their associated angular momenta.

We consider an agent-based model of $N$ interacting point-like particles 
moving in two dimensions.  The state of each agent $i$ at a time step $t$
is described by its position $\mathbf{x}_i(t)$ and velocity $\mathbf{v}_i(t)$.
We define the velocity of an agent to be its displacement in a unit time
step, and which therefore has the dimension of length.
The dynamics of the system is governed by the following update rule:
at each time step $t$, an agent $i$ interacts  with a randomly chosen agent $j$
with a specified probability $p(\mathbf{x}_j(t),\mathbf{v}_j(t)|\mathbf{x}_i(t),\mathbf{v}_i(t))$,
defined later, leading to a change in its velocity. If it does not find any
agent to interact with, it instead moves a distance $|\mathbf{v}_i(t)|$ in a
random direction. The velocity $\mathbf{v}_i(t)$ and position $\mathbf{x}_i(t)$
are updated as
\begin{subequations}
\begin{align} 
\label{vel-update}
\mathbf{v}_i(t+1) &= \mathbf{v}_i(t) + \mathbf{a}_i(t)\,,\\
\label{pt-update}
\mathbf{x}_i(t+1) &= \mathbf{x}_i(t) + \mathbf{v}_i(t+1)\,.
\end{align}
\label{eq-update}
\end{subequations}
Here $\mathbf{a}_i(t)$ is the acceleration of the agent, and is given by
\begin{align}
  \mathbf{a}_i(t) &=\left\{ \begin{array}{ll}
      -\mathbf{v}_i(t) + |\mathbf{v}_i(t)|\,\hat{\bm{\eta}}, &\text{if~} \Omega_i = \emptyset, \\
      \alpha[\mathbf{v}_j(t)-\mathbf{v}_i(t) +  f(\mathbf{v}_j(t)+\mathbf{v}_i(t))], &\text{otherwise},
                           \end{array}
    \right.
\label{stoch-accel}
\end{align}
where $\Omega_i$ is the set of all agents with which agent $i$ may interact
with, the coefficient $\alpha < 1$ is the strength of interaction, and
$\hat{\bm{\eta}}$ is a chosen from a uniform random distribution
of vectors on the unit circle.
The initial condition is specified as $\mathbf{x}_i(0)=\mathbf{x}_i^{0}$ and
$\mathbf{v}_i(0)=\mathbf{v}_i^{0}$ for all $i=1,2,\ldots N$.

We note from \eq{vel-update} that when $\Omega_i \neq \emptyset$, the velocity update is dependent on the
randomly chosen agent $j$. The linear term $\alpha(\mathbf{v}_j-\mathbf{v}_i)$
in \eq{stoch-accel} describes an alignment interaction, while the nonlinear
term $f(\mathbf{v}_j+\mathbf{v}_i)$ keeps the velocity close to a critical value
$\mathbf{v}_c$, i.e. it ensures that the flock maintains a constant average speed. 
Assuming $|\mathbf{v}_c|=1$, we consider 
$f(\mathbf{v}) := \mathbf{v}(1-|\mathbf{v}|)/(1+|\mathbf{v}|^\beta)$ with $\beta =3$
(see Supplementary Information for a more detailed discussion).
\begin{figure}[!t]
\includegraphics{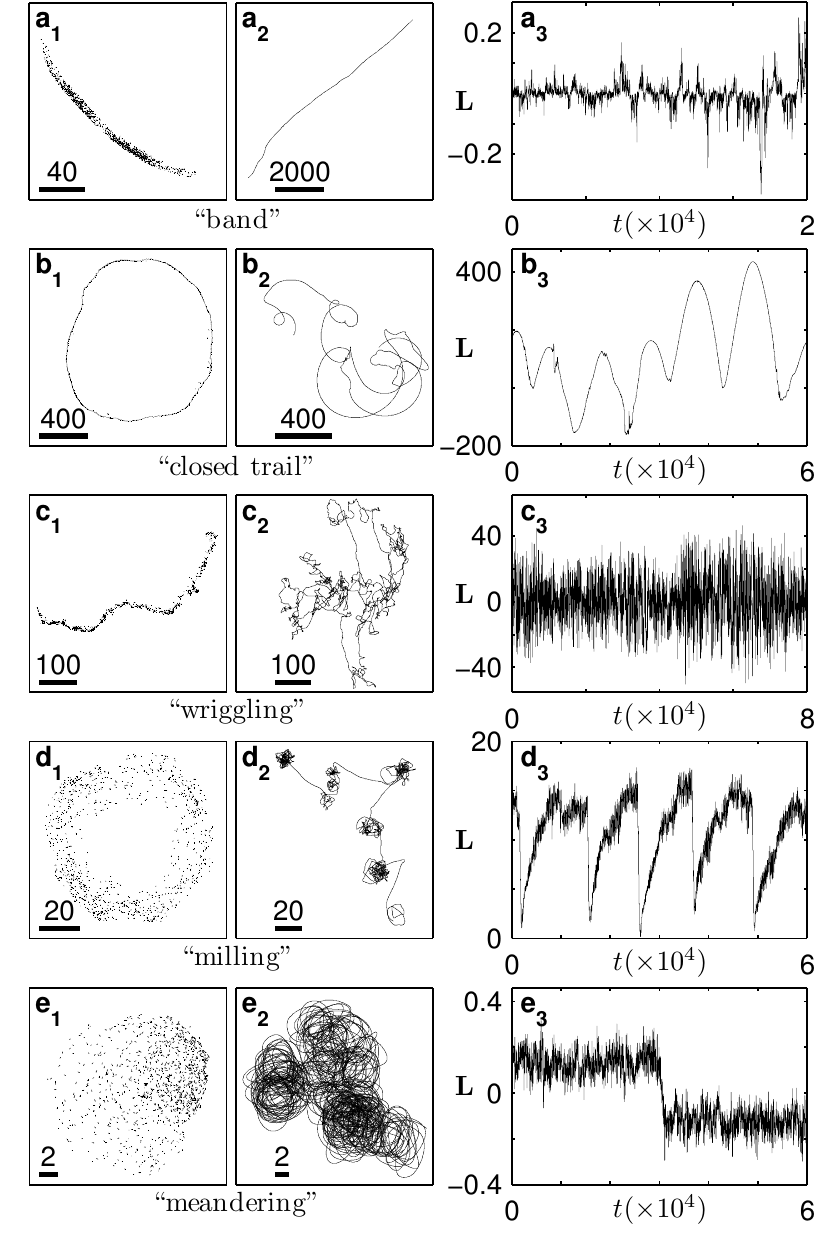}
\caption{Examples of the spatially contiguous dynamical flocking patterns
exhibited by the model for a system of $N=10^{3}$ agents. In each row the
left panel displays a snapshot of the flock, the right panel displays the angular
momentum per particle $\mathbf{L}$ over a duration of time, and the middle
panel displays the corresponding trajectory of the center of mass of the flock
$\bar{\mathbf{x}}(t)$ over the same duration. (a$_{1}$-a$_{3}$) Agents moving
in a band for the case $\sigma=6$, $\theta_{\max}=90$ and $\alpha=0.1$.
(b$_{1}$-b$_{3}$) Agents moving in a wriggling pattern for the case
$\sigma=5$, $\theta_{\max}=40$ and $\alpha=0.8$. (c$_{1}$-c$_{3}$) Agents
moving in a closed trail for the case $\sigma=3$, $\theta_{\max}=50$ and
$\alpha=0.1$. (d$_{1}$-d$_{3}$) Agents moving in a milling pattern for the
case $\sigma=1$, $\theta_{\max}=20$ and $\alpha=0.025$.
(e$_{1}$-e$_{3}$) Agents moving in a flock with a meandering center of mass
for the case $\sigma=3$, $\theta_{\max}=15$ and $\alpha=0.02$. The
numbered solid bars in the left and middle panels of every row provides a
measure of spatial distance in each case.
}
\label{fig1}
\end{figure}

%
For our current investigation, we assume that every agent $i$ has a field
of view, symmetric around its direction of motion, that is delimited by a
maximum bearing angle $\theta_{\max}$. 
The probability $p(\mathbf{x}_{j},\mathbf{v}_{j}|\mathbf{x}_{i},\mathbf{v}_{i})$ that an agent $i$
interacts with an agent $j \in \Omega_i$ may be specified in terms of weights $\omega_{i,j}$. 
We assume that a given agent mostly interacts with agents separated from it by
an optimal interaction length, and that the probability that it randomly selects an
agent lying very close to, or very far away from itself is negligible.
With these properties in mind we assume the following weight function
\begin{equation}
\omega_{i,j}=|\mathbf{x}_i-\mathbf{x}_j|\,e^{-\frac{|\mathbf{x}_i-\mathbf{x}_j|^2}{2\sigma^2}}\left(1-{\theta_{i,j}^2}/{\theta_{\max}^2}\right)\,,
\label{eq:transitionrate}
\end{equation}
if $\theta_{i,j} \leq \theta_{\max}$ and $\omega_{i,j}=0$ for $\theta_{i,j} > \theta_{\max}$,
where $\sigma$ is the mean interaction length
and $\theta_{i,j}$ is the angle between the velocity $\mathbf{v}_i$ and the vector $\mathbf{x}_j-\mathbf{x}_i$.
Given this weight function, the probability can be written as 
$p(\mathbf{x}_{j},\mathbf{v}_{j}|\mathbf{x}_{i},\mathbf{v}_{i}) = \omega_{i,j}/\sum_{k \in \Omega_i} \omega_{ik}$.

In the limiting case $\theta_{\max}=\pi$, there are no random rotations as,
by definition, we would have $\Omega_{i} \neq \emptyset \;\forall\; i$.
 In this situation any initial randomness will eventually get redistributed over the
whole population, and it is expected that the velocities will converge to that of
the initial mean velocity. Furthermore, here an agent $i$ has the highest
likelihood to align with any neighbour $j$ that approximately lies at a distance
$|\mathbf{x}_{i}-\mathbf{x}_{j}|=\sigma$ (i.e. where $\omega_{i,j}$ is at its
maximum). Hence, in our simulations we assume that the initial positions
$\mathbf{x}_i^{0}$ are selected randomly over a small region of size
$\sim \mathcal{O}(\sigma)$ and velocities $\mathbf{v}_i^{0}$ are chosen from a
uniform distribution.
\begin{figure}[!t]
\includegraphics{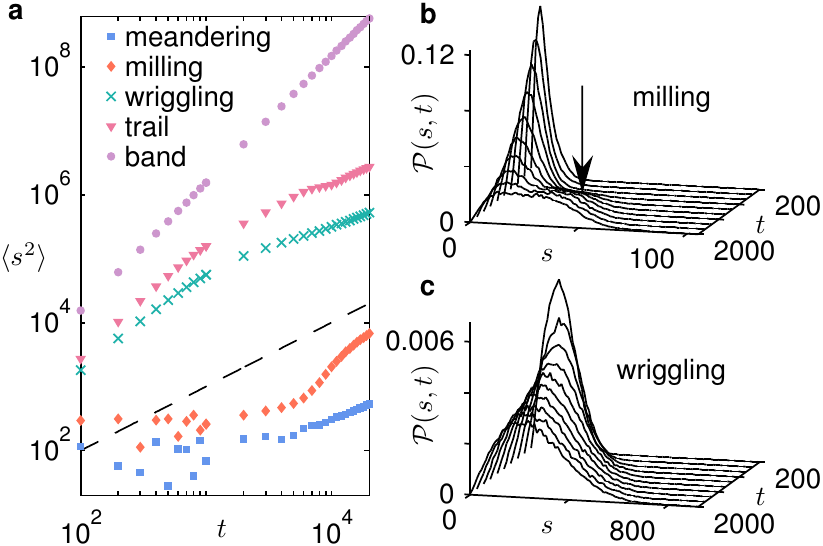}
\caption{Statistics of the center of mass trajectories (a) Time-dependence of the
average mean-squared displacement (MSD) of the center of mass $\langle s^2 \rangle$,
calculated over $10^4$ trials, for each of the five sets of parameter values
considered in Fig.~\ref{fig1}. The dashed line, shown for reference, indicates the
MSD for the case of normal diffusion. (b-c) The probability distribution function
$\mathpzc{P}(s,t)$, calculated over $5\times10^4$ trials, shown over a
range of displacements $s$ and time $t$ for the cases (b) $\sigma=1$,
$\theta_{\max}=20$, $\alpha=0.025$ (a milling pattern), and (c) $\sigma=5$,
$\theta_{\max}=40$, $\alpha=0.8$ (a wriggling pattern).
The arrow in panel (b) indicates a large excursion.
}
\label{fig2}
\end{figure}
%
Upon varying the interaction strength $\alpha$, mean interaction length $\sigma$ and
the maximum bearing angle $\theta_{\rm max}$ over a range of values for a system
of $N=10^{3}$ agents, we find that the model exhibits a wide range of patterns
(see \fig{fig1}). From our numerical simulations, we find that the resulting patterns can sustain their
cohesiveness over a very long period of time ($t \gtrsim 10^{6}$ steps). These
observed patterns include an extended band-like flock that can move ballistically
for long durations (\fig{fig1}(a)), a spatially extended wriggling pattern
(\fig{fig1}(b)), a very large and narrow closed trail pattern (\fig{fig1}(c)), a
flock that exhibits a milling, or vortex-like, pattern (\fig{fig1}(d)), and a
flock with a meandering center of mass, and rotating profile, that remains
confined to a small region of space (\fig{fig1}(e)). Movies of the patterns
displayed in \fig{fig1}(b$_{1}$-e$_{1}$) are included as Supplementary
Information. Furthermore, in addition to the patterns displayed in \fig{fig1}, this system
can exhibit multiple interacting clusters. To illustrate this we have plotted in \fig{fig1}(a$_{3}$-e$_{3}$) the
temporal variation of the angular momentum per particle,
$\mathbf{L} = N^{-1}\sum_i (\mathbf{x}_i-\bar{\mathbf{x}}) \times \mathbf{v}_i$ 
for the corresponding flocking patterns, where
$\bar{\mathbf{x}}(t)=N^{-1}\sum_{i} \mathbf{x}_{i}(t)$ is the center of mass of
the flock. We observe that this quantity exhibits remarkably distinct temporal
profiles for each of the displayed patterns, and captures the spontaneous
switching/reversal in the direction of rotation of the flock, which
manifests as a change in the sign of $\mathbf{L}$.
%
\begin{figure}[!t]
\includegraphics{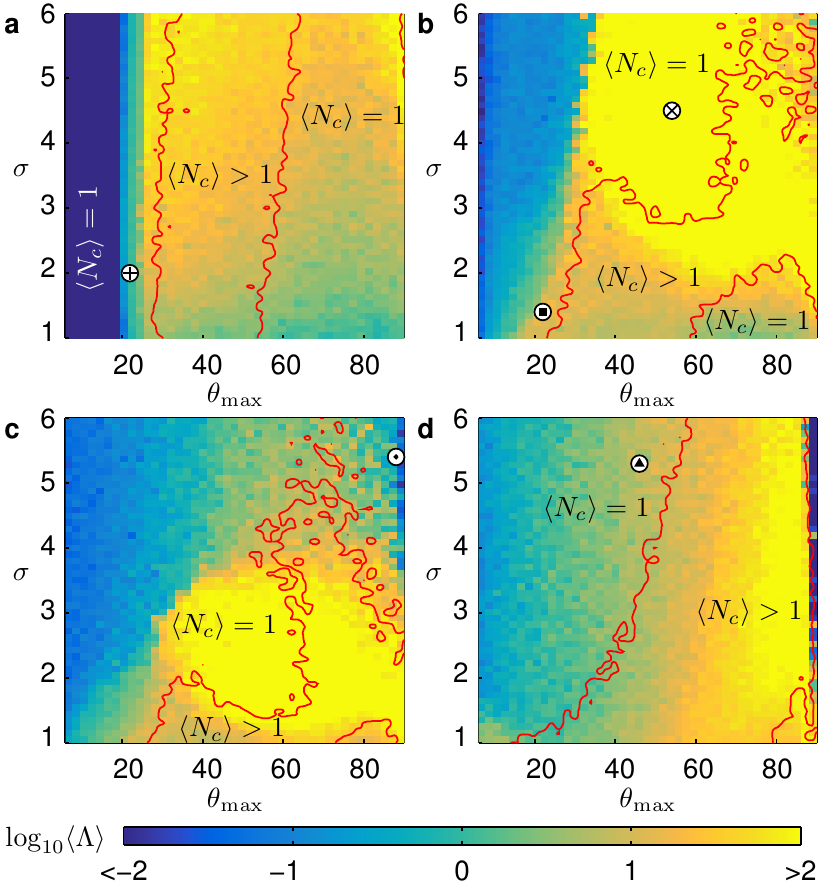}
\caption{
Parameter space diagrams obtained using the cluster-finding algorithm 
described in the text. The ensemble-averaged quantities
$\langle N_{c}\rangle$ and $\langle \mathbf{\Lambda}\rangle$ are
computed over a range of values of the mean interaction length
$\sigma$, interaction strength $\alpha$ and the maximum bearing angle $\theta_{\max}$,
and are averaged over $10$ trials. The four panels correspond to (a) $\alpha=0.01$,
(b) $\alpha=0.05$, (c) $\alpha=0.1$ and (d) $\alpha=0.5$.
In each, we display (in log-scale) the dependence of the average angular
momentum of the flock $\langle \mathbf{\Lambda}\rangle$ on system
parameters, along with contour lines that
demarcate the regimes where the flock is characterized by a single cluster
($\langle N_{c}\rangle=1$) and multiple clusters ($\langle N_{c}\rangle>1$).
The black markers within white circles in each panel indicate locations
in the parameter space
where we observe a meandering pattern (a: plus sign), a milling pattern
(b: filled square), a closed trail (b: cross), a band pattern (c: filled circle) and a
wriggling pattern (d: filled triangle). A more detailed exploration of the parameter
space, with snapshots of the patterns obtained, is provided in the
Supplementary Information.
}
\label{fig3}
\end{figure}

In Fig.~\ref{fig1}(a$_{2}$-e$_{2}$), the trajectories of the center
of mass of the flock, $\bar{\mathbf{x}}(t)$, illustrate the diversity of collective
dynamics that this model is capable of exhibiting. These range from near-ballistic
motion in the case of the band-like patterns (\fig{fig1}(a2)) to winding behaviour
with occasional long excursions, similar to that of a
correlated random walk, in the
case of the milling pattern (\fig{fig1}(e2)). To discern the macroscopic features of
these trajectories, we discard an initial transient period of duration
$t_{0}=10^{3}$ and compute the probability distribution function $\mathpzc{P}(s,t)$,
where $s = |\bar{\mathbf{x}}(t) - \bar{\mathbf{x}}(t_{0})|$, and the mean square
displacement (MSD) of the center of mass, $\langle s^2\rangle$. While the trail and
wriggling patterns show a superdiffusive behaviour at small time scales, they appear to
converge to normal diffusion $\langle s^2\rangle \sim t$ asymptotically (cf. dashed
line in \fig{fig2}(a)). In contrast, the milling and the meandering patterns are
initially subdiffusive and asymptotically converge to normal diffusion, while the
band pattern is superdiffusive at all times. The probability density function
$\mathpzc{P}(s,t)$ for the milling and the wriggling patterns are shown in \fig{fig2}(b)
and (c). We find that the patterns show a qualitatively similar decay of
$\mathpzc{P}(s,t)$ at small times. However, as indicated by an arrow in
\fig{fig2}(b), the center of mass of the milling pattern exhibits a higher
probability of large excursions at later times, which corresponds to
intervals where rotation ceases due to an internal
reorganization of the flock.

It is apparent from the breadth of complexity of the observed flocking
patterns that simple order parameters, such as the mean velocity of the
flock, would be insufficient to characterize the dynamics of the model.
While a non-zero mean velocity, corresponding to ordered motion, may
indicate the existence of the band pattern,
a zero mean velocity may either correspond to diffusive 
randomly moving agents or to an ordered rotating swirl.
Furthermore, we find that the flock
may be characterized by several clusters for certain choices of the system
parameters. Hence, we would
require a set of order parameters that could more accurately distinguish
between the wide array of flocking patterns observed in our simulations

To this end, we classify the patterns in terms of the number of distinct
(contiguous) clusters and their associated angular momenta at a given
time, through a cluster-finding algorithm. This procedure, which we
rigorously detail in the Supplementary Information, is outlined as follows.
We define the resolution length $R=\lambda R_{\max}$, where
$0 < \lambda \leq 1$, and $R_{\max}$ is the maximum separation between any
two particles in the flock at time $t$. At the length scale $R_{\max}$ the
system can be viewed as comprising a single cluster that encompasses the
entire flock. For the chosen length scale $R$, we first compute
$r_{i,j}=|\mathbf{x}_i(t)-\mathbf{x}_j(t)|$ for all $i$,$j \neq i$, and
group the agents into distinct clusters such that a pair of agents ($i,j$)
in any given cluster satisfies the condition $r_{i,j} \leq R$. Next, we
regroup the agents such that if $r_{i,j} \leq R$ and $r_{j,k} \leq R$ but
$r_{i,k} > R$ then the agents $i,~j,$ and $k$ are assumed to belong to the
same cluster. The resolution length $R$ hence provides a lower bound on the
spatial separation of any pair of detected clusters. Once the individual
clusters $c_{i}$ (of size $N_{i}$) have been determined, we define $N_{c}$
to be the minimum number of clusters whose collective population exceeds
$90\%$ of $N$, i.e.
$N_{c}=\min\left\{n: 0.9\,N \leq \sum_{i=1}^{n}N_{i}, 1\leq n\leq N \right\}$.
The center of mass of a cluster $c_i$ is defined as
$\bar{\mathbf{x}}_i=N_i^{-1}\sum_{j \in c_i}\mathbf{x}_j$, and the
corresponding angular momentum about the center of mass is
$\mathbf{L}_{i}=N_i^{-1}\sum_{j\in c_i}(\mathbf{x}_j-\bar{\mathbf{x}}_i)\times \mathbf{v}_j$.
We then compute the quantity $\Lambda = N_c^{-1}\sum_{i=1}^{N_c} |\mathbf{L}_i|$,
where the absolute value sign takes into account the fact that
the flock may contain clusters that swirl in
opposite directions. In our simulations we have used $\lambda=2^{-4}$,
and find that a small variation $R \pm \delta$, where $\delta\in(0,R/2)$,
does not affect the classification of the patterns. Note that in the limit
$\lambda \rightarrow 0$ we would, by definition, find $N$ clusters that
each comprise a single agent.

In \fig{fig3} we display a parameter space diagram that 
classifies the flocking patterns in terms of two ensemble averaged
quantities, namely angular momentum 
$\langle \Lambda \rangle$ and the number of clusters  $\langle N_c \rangle$, over
a range of values of $\sigma$, $\theta_{\max}$ and $\alpha$.
The contour lines demarcate regimes where the flocking pattern is characterized
by a single ($\langle N_{c}\rangle=1$) and multiple clusters ($\langle N_{c}\rangle>1$).
A general observation from \fig{fig3} is that at low values of $\theta_{\rm max}$, the mean angular momentum 
is very low, regardless of $\sigma$ or $\alpha$ and that the corresponding patterns are characterized by a single diffusive cluster.
Such cohesive but highly disordered flocking behaviour has been reported
earlier in the context of midge swarming patterns~\cite{Okubo1986}.
Patterns with very high angular momentum, which typically correspond to single
or multiple closed trails, are observed for larger values of $\alpha$.
For $\alpha = 0.01$ we observe multiple clusters over
an intermediate range of values of $\theta_{\rm max}$.
Multiple clusters are also observed for larger values of $\alpha$, although
the regimes where they occur exhibit a more complex dependence on $\theta_{\rm max}$.
Several snapshots of the collective patterns obtained
over the entire range of parameter values displayed in \fig{fig3} are
presented in the Supplementary Information.

A crucial feature of our model is that the stochasticity is maximum at the
edges of the flock, while the stochastic velocity alignments in the interior
of the flock gives rise to comparatively ordered behaviour through a process
of self-organization. In addition to facilitating cohesion, this may help
explain the apparent symmetry of several of the patterns (c.f. milling,
meandering and closed trails), as flocks with relatively smoother boundaries
have much lower stochasticity overall. In other words, the overall
stochasticity reduces through a minimization of surface area. In this regard,
the existence of the wriggling pattern, which has a rougher boundary, is due
to the fact that the stochasticity at the edge is reduced for larger values
of $\sigma$. These results are intriguing in light of recent observations
that the boundary of a flock plays an important role in its emergent dynamical
properties~\cite{Cavagna2013a}. Additionally, we note that as the alignment
probability in our model is dependent on $\theta_{\max}$, there is an inherent
spatial anisotropy in the stochastic interactions. Specifically, for
$\theta_{\max}<90$ agents do not interact with neighbours that lie directly
behind them. This may relate to the emergence of milling patterns in our model,
as previous flocking models that reported such patterns have typically
incorporated such a ``blind zone''
for agents~\cite{Couzin2002,Lukeman2008,Pearce2014,Costanzo2018}. This pattern has been
observed in diverse contexts across the natural world
~\cite{Lukeman2008,Lopez2012,Tunstrom2013,Sendova-Franks2018}, including fish schools
and ant mills. Furthermore,
it can be seen that $\Omega_{i}$ is not invariant under the transformation
$\mathbf{v}_i \rightarrow -\mathbf{v}_i$, as a consequence of the inherent
anisotropy of the field of view, which hence breaks the time-reversal symmetry.
However, such a transformation will not affect the nature of the pattern
at the scale of the entire flock. Finally, there remain intriguing questions
related to the nature of phase transitions that this system may exhibit, as
well as the role of system size. However, we would like to emphasize that
the nature of inter-particle interactions in this model suggests that the nature
of the emergent behaviour would depend more on the density than on the total
number of particles in the system.

In conclusion, our model provides a mechanism through which stochasticity
arises intrinsically from the interactions between agents. This framework
can, in principle, be generalized to the case of stochastic many-body
interactions. In addition, our cluster-finding method characterizes the
rich dynamical patterns observed in terms the number of spatially
distinct clusters of the flock and their angular momenta.
As this algorithm is independent of the details of the flocking mechanism, it
may help provide additional insights into other flocking systems, both
theoretical and experimental.
Furthermore, the
model proposed here could be extended to describe situations of pursuit
and evasion in predator-prey systems~\cite{Romanczuk2009}, as well as
incorporate the role of social hierarchy in flocks~\cite{Nagy2010,Petit2015,Lopez2018}.
%

\begin{acknowledgments}
We would like to thank
Abhijit Chakraborty, Niraj Kumar, V. Sasidevan and Gautam Menon
for helpful discussions. SNM is supported by the IMSc Complex Systems
Project ($12^{\rm th}$ Plan). The simulations and computations required
for this work were supported by the Institute of Mathematical Science's
High Performance Computing facility (hpc.imsc.res.in) [nandadevi],
which is partially funded by DST.
\end{acknowledgments}



\onecolumngrid

\setcounter{figure}{0}
\renewcommand\thefigure{S\arabic{figure}}  
\renewcommand\thetable{S\arabic{table}}

\vspace{12cm}

\section*{\large {\bf SUPPLEMENTARY INFORMATION}}
\vspace{2cm}


\section*{Contents}

\begin{enumerate}
	\item Schematic of an agent's field of view
	\item Algorithm for computing the number of clusters
	\item Detailed explanation of the nonlinear term in the model
	\item Snapshots of flocking patterns observed over a range of parameter values
	\item Description of the movies
\end{enumerate}

\renewcommand\thepage{S\arabic{page}}
\setcounter{page}{1}

\section*{Schematic of an agent's field of view}

\begin{figure}[!h]
\centering
\includegraphics{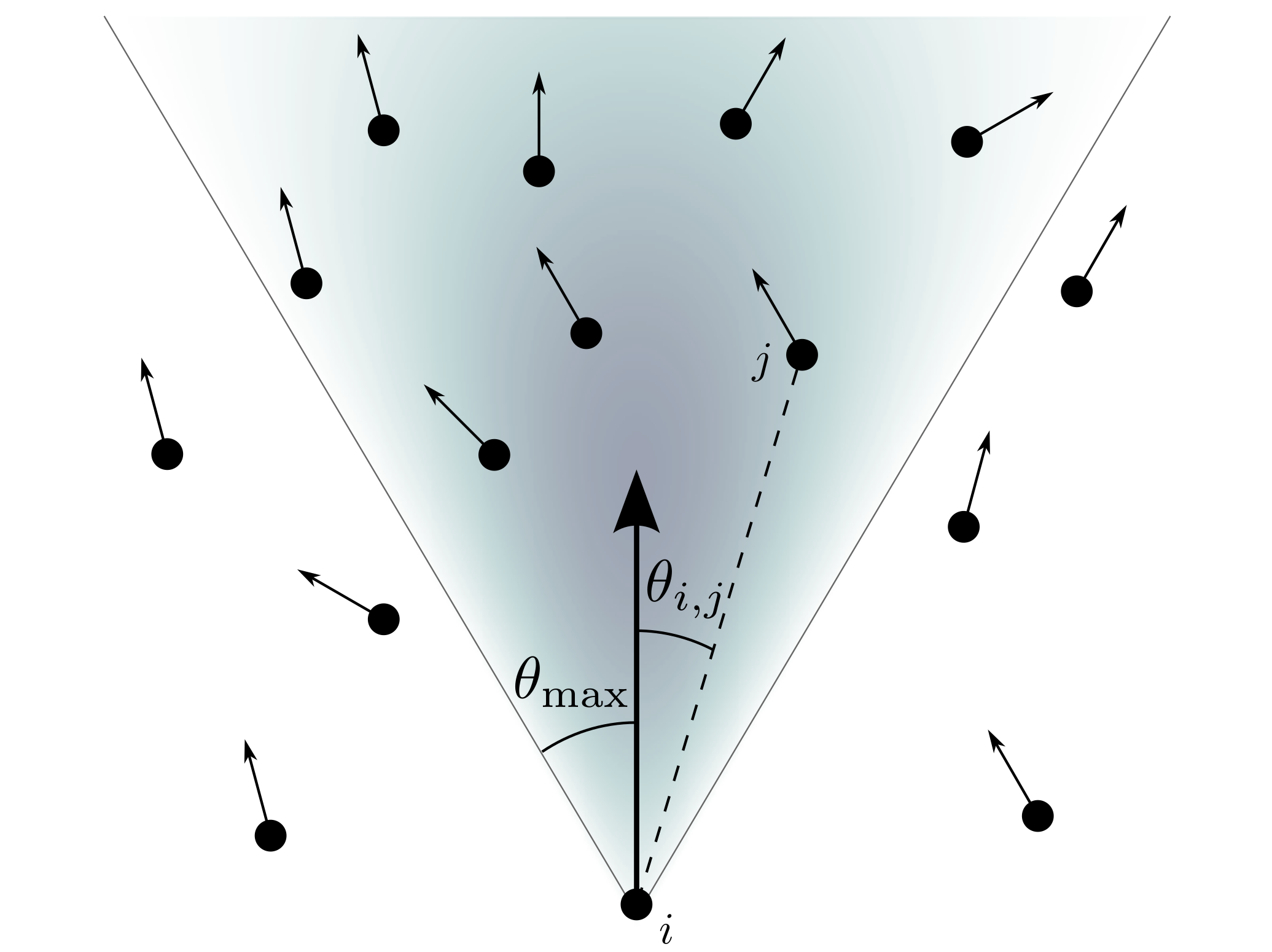}
\caption{
Schematic of the field of view of an agent $i$ that picks an agent $j$ lying within this field
of view. The intensity of colour in a given region is related to the the probability with which
agent $i$ chooses an agent that lies in that region. Each agent has the highest probability of
interacting with agents that lie at a distance $\sigma$ along its direction of motion.
Similarly, the intensity reduces as the angle $\theta_{i,j}$ between the velocity of $i$ and
the line connecting the agents approaches the maximum bearing angle $\theta_{\rm max}$. Thus,
an agent $i$ is most likely to align with an agent that is near its direct line of sight, and
which is separated by a distance of around $\sigma$.
}
\label{sch}
\end{figure}

The field of view of agent $i$ is illustrated in Fig.~\ref{sch}. At each iteration, agent $i$ 
attempts to select an agent that lies within its field of view, which is delimited by a maximum
bearing angle $\theta_{\text{max}}$, for the purposes of an alignment interaction. An agent $j$
within this field of view is picked by $i$ with a probability that is related to the distance
between them, as well as the angle between the velocity of $i$ and the line connecting the two
agents. If the field of view of agent $i$ is empty, it performs a random rotation.

\clearpage

\section*{Algorithm for computing the number of clusters}

At any specified time instant, the maximum possible distance between a pair of agents in the flock
is denoted by
\[
R_{\max}=\max\left(|\mathbf{x}_i(t)-\mathbf{x}_j(t)|\right)\,
\forall\,
i,j\in [1,N]\,.
\]

\begin{figure}[!ht]
\centering
\includegraphics[width=\textwidth]{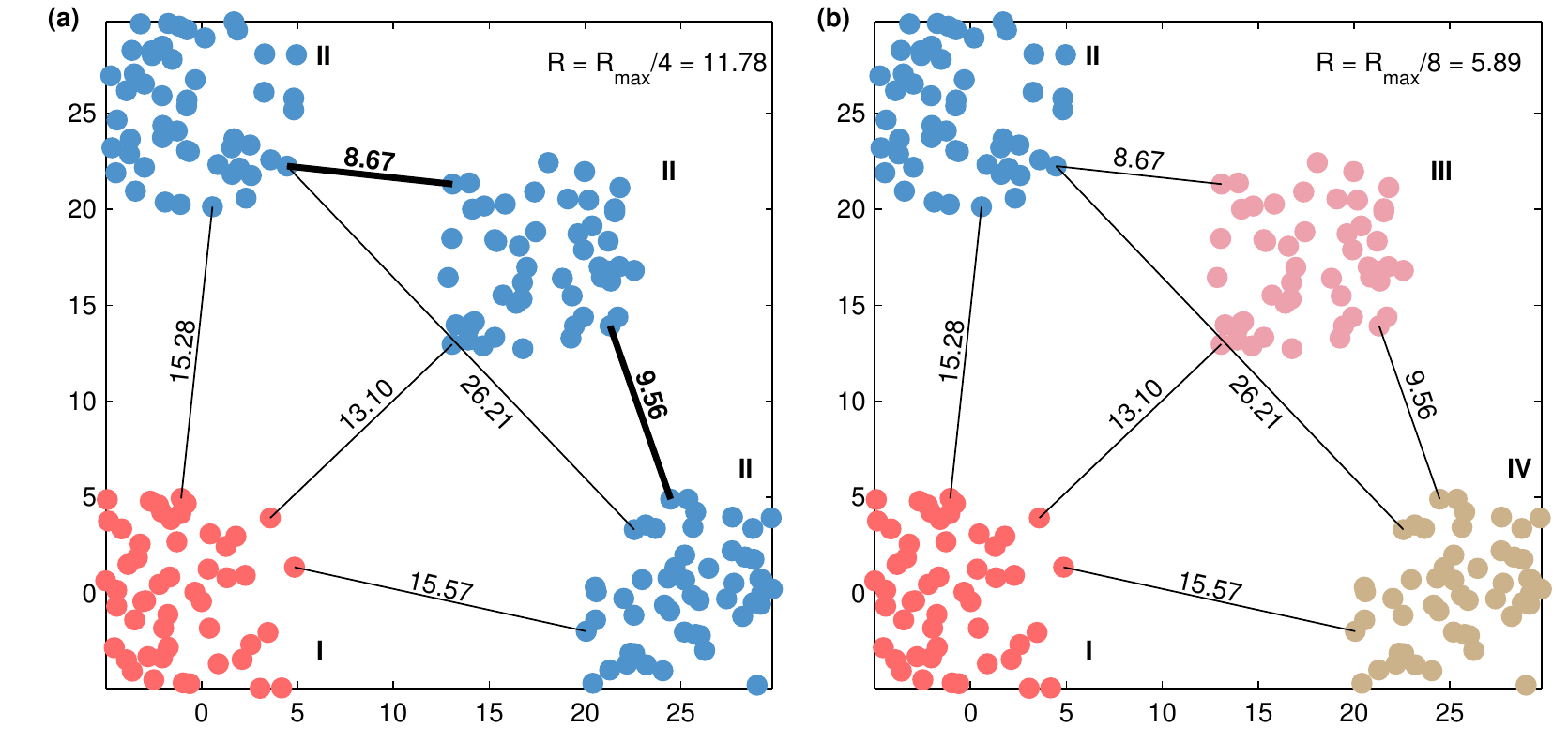}
\caption{
A demonstration of the cluster-finding algorithm.
We choose resolution lengths (a) $R=R_{\max}/4$, and (b) $R=R_{\max}/8$. The lines connect
the closest agents in each pair of clusters, and the corresponding numerical value denotes
the distance between these agents. The bold lines and numbers in panel (a) indicate that
the corresponding clusters are categorized as being part of the same cluster (II). In panel
(b) four clusters (I-IV) are obtained since all of them are separated by a distance $>R$.
}
\label{algo_example}
\end{figure}

We set the resolution length $R = \lambda\, R_{\max}$ by choosing a value of $\lambda$ in the
range $0 < \lambda \leq 1$. Each agent $i=1,2,\ldots,N$ is assigned a label $g_{i}$ which is
associated with an integer value that specifies the cluster to which the agent belongs to. The
cluster-finding algorithm involves determining the number of distinct clusters $N_{c}$ of size
$\geq R$. The label of each agent $i$ thus lies in the range
$g_{\min}(=1) \leq g_i \leq g_{\max}(=N_{c})$.\\
~\\
{\bf{Summary of the variables used:}}\\
~\\
\begin{tabular}{rl}
$N$ : 		 & Total number of agents in the system,\\
$N_c$ :		 & Total number of clusters found using the algorithm,\\
$R$ :		 & Resolution length of the flock (defined above),\\
$g_i$ :		 & Label associated with each cluster,\\
$b_i$, $c$ : & Boolean variables,\\
$g_{\min}$ : & Minimum value of the array $g$,\\
$g_{\max}$ : & Maximum value of the array $g$.
\end{tabular}
~\\
~\\
~\\
{\bf{Pseudocode of the algorithm:}}\\
~\\
The algorithm is outlined in the following pseudocode. Comments appear in blue italicised text.
\begin{tabbing}
\hspace*{1.0cm}\=\hspace*{1.0cm}\=\hspace*{1.0cm}\=\hspace*{1.0cm}\=\hspace*{1.0cm}\= \kill
\TT{Initalize:} $g_{\max} = 0$, $g_{\min} = 0$, \TT{and} $g_i = 0$ \TT{for all} $i=1,2,\ldots,N$.\\
\TT{For} $i=1,2,\ldots,N$\\
\>\CMT{If agent $i$ has not been assigned a label, we label it as one plus the maximum value of
the array $g$.}\\
\>\TT{If} $g_i = 0$ \TT{Then} $g_i=\max \{ g_{i'}, i'=1,2,\ldots,N \}+1$.\\
\>\CMT{The variable $b$ marks all the agents in the current assignment.}\\
\>\TT{Initalize:} $b_j = 0$ \TT{for all} $j=1,2,\ldots,N$.\\
\>\CMT{Find all agents $j$ that are at a distance $\leq R$ from agent $i$ and assign $j$ with
the same label as $i$. }\\
\>\TT{For} $j=1,2,\ldots,N$\\
\>\>\TT{If} $|\mathbf{x}_i(t)-\mathbf{x}_j(t)|<R$ \TT{Then}\\
\>\>\>\TT{If} $g_j = 0$ \TT{Then} $g_j=g_i$.\\
\>\>\>$b_j=1$.\\
\>\>\TT{End}\\
\>\TT{End}\\
\>\TT{Initalize:} $g_{\min}=g_i$.\\
\>\CMT{Consider all the marked agents, i.e. all agents $j$ for which $b_j=1$.}\\
\>\CMT{We find the minimum value of $g_j$ and assign it to $g_{\min}$}\\
\>\TT{For} $j=1,2,\ldots,N$\\
\>\>\TT{If} $b_j = 1$ \TT{Then}\\
\>\>\>\TT{If} $g_j \leq g_{\min}$ \TT{Then} $g_{\min}=g_j$.\\
\>\TT{End}\\
\>\TT{For} $j=1,2,\ldots,N$\\ 
\>\>\CMT{We assign the minimum value of the array $g$ to all the marked agents.}\\
\>\>\TT{If} $b_j = 1$ \TT{Then}\\
\>\>\>\TT{For} $k=1,2,\ldots,N$\\
\>\>\>\>\TT{If} $g_k = g_j$ \TT{and} $k \neq j$ \TT{Then} $g_k=g_{\min}$.\\
\>\>\>\TT{End}\\
\>\>\>$g_j=g_{\min}$.\\
\>\>\TT{End}\\
\>\TT{End}\\
\TT{End}\\
~\\
\TT{Compute:} $g_{\max} = \max \{ g_{i'}, i'=1,2,\ldots,N \}$.\\
~\\
\CMT{If more than one cluster exists, we relabel them so as to remove the value zero.}\\
\TT{If} $g_{\max} > 1$ \TT{Then}\\
\>\TT{For} $i=(g_{\max}-1),~(g_{\max}-2),\ldots,1$\\
\>\>\TT{Set:} $c=0$\\
\>\>\TT{For} $j=1,2,\ldots,N$\\
\>\>\>\TT{If} $g_j = i$ \TT{Then} $c=1$ \TT{and Exit}.\\
\>\>\TT{End}\\
\>\>\CMT{Fix gaps in the label numbers to ensure that the final set is contiguous}\\
\>\>\TT{If} $c=0$ \TT{Then}\\ 
\>\>\>\TT{For} $j=1,2,\ldots,N$\\
\>\>\>\>\TT{For} $k=i+1,\ldots,g_{\max}$\\
\>\>\>\>\>\TT{If} $g_j = k$ \TT{Then} $g_j=k-1$.\\
\>\>\>\>\TT{End}\\
\>\>\>\TT{End}\\
\>\>\TT{End}\\
\>\TT{End}\\
\TT{End}\\
~\\
\TT{Compute:} $g_{\min} = \min \{ g_{i'}, i'=1,2,\ldots,N \}$, $g_{\max} = \max \{ g_{i'}, i'=1,2,\ldots,N \}$.
\end{tabbing}
Once each $g_i$ has been relabelled, the number of agents in each cluster $i$ is simply the
number of agents that are labelled $g_{i}$, and the total number of clusters at the chosen
resolution length $N_{c}=g_{\max}$.\\
~\\
{\bf{Demonstration of cluster-finding algorithm at different resolution lengths:}}\\
~\\
In the following example, we present an implementation of this cluster-finding algorithm at
two different resolution lengths, $R$. As displayed in Fig.~\ref{algo_example}, we consider
four clusters of agents. Each cluster consists of $50$ agents whose coordinates are chosen
randomly within a $10\times 10$ square centered at the coordinates $(0,0)$, $(0,25)$, $(25,0)$,
and $(25/\sqrt{2},25/\sqrt{2})$.

\clearpage

Upon running our cluster-finding algorithm on this flock, we find that the maximum separation
between any pair of agents is $R_{\max}=47.13$. For the choices $\lambda=1/4, 1/8$, we find
$R=R_{\max}/4=11.78$ and $R=R_{\max}/8=5.89$. In the displayed realization
(Fig.~\ref{algo_example}), we find that the minimum distance between agents in the lower left
and upper right clusters is $13.1$. Hence, at resolution length $R=11.78$ these two clusters
are categorized as being distinct. In contrast the minimum distances between the agents in
upper right cluster and those in the remaining clusters are less than $11.78$ and hence they
are categorized as being part of the same cluster. Thus, as displayed in
Fig.~\ref{algo_example}(a), at resolution length $R=11.78$ we find just two distinct clusters
I \& II (coloured red and blue).\\

For the case where a resolution length $R=R_{\max}/8=5.89$ is used, we find that since all
four clusters are separated by a value greater than $R$ they are categorized are being
distinct. Thus, our method obtains four distinct clusters (I-IV) at this resolution length,
as displayed in Fig.~\ref{algo_example}(b) where each cluster is coloured distinctly.


\section*{Detailed explanation of the nonlinear term in the model}

In Eq. (2) of the main text, we introduce a nonlinear term
$f(\mathbf{v}_j+\mathbf{v}_i)$, where $\mathbf{v}_i$ and $\mathbf{v}_j$
are respectively the velocities of agents $i$ and $j$. For the purpose of
the current investigation, we consider the functional form
$f(\mathbf{v}) := \mathbf{v}(1-|\mathbf{v}|)/(1+|\mathbf{v}|^\beta)$ with $\beta =3$.
Note that if the field of view of agent $i$ is nonempty, i.e. $\Omega_{i}\neq\emptyset$,
its velocity at time step $t+1$ is:
\[
\mathbf{v}_i(t+1) = \mathbf{v}_i(t) +  \alpha[\mathbf{v}_j(t)-\mathbf{v}_i(t) +  f(\mathbf{v}_j(t)+\mathbf{v}_i(t))]
\]
For the functional form that we consider, we see that
$f(\mathbf{v}_j+\mathbf{v}_i)$ vanishes
at $|\mathbf{v}_j+\mathbf{v}_i| = 0$, $1$ and infinity, which implies that the
velocity
$\mathbf{v}_i(t+1) \simeq \mathbf{v}_i(t) + \alpha (\mathbf{v}_j(t)-\mathbf{v}_i(t))$
near these values. The case $|\mathbf{v}_j+\mathbf{v}_i| = 0$ corresponds to a
situation where the velocities of particles $i$ and $j$ have identical magnitudes
and opposite directions. In this scenario, the resulting velocity update effectively
prevents a direct collision.

To understand the case $|\mathbf{v}_j+\mathbf{v}_i| = 1$, let us
assume that $|\mathbf{v}_i+\mathbf{v}_j| = 1 + \epsilon$,
where $|\epsilon| \ll 1$. In this situation, we see that
\[
f(\mathbf{v}_i+\mathbf{v}_j) = \frac{(\mathbf{v}_i+\mathbf{v}_j)(1-(1+\epsilon))}{1+(1+\epsilon)^3} \simeq  \frac{-\epsilon(\mathbf{v}_i+\mathbf{v}_j)}{2}\,,
\]
Substituting this expression into Eq. (2) of the main text, we find that the acceleration is
\[
\mathbf{a}_i \simeq \alpha\left(1 - \frac{\epsilon}{2}\right)\mathbf{v}_j -
                    \alpha\left(1 + \frac{\epsilon}{2}\right)\mathbf{v}_i\,
\]
Using the velocity update expression from Eq. (1) of the main text,
we see that $|\mathbf{v}_i|_{\epsilon \neq 0} < |\mathbf{v}_i|_{\epsilon=0}$ if 
$\epsilon >0$, and $|\mathbf{v}_i|_{\epsilon \neq 0} > |\mathbf{v}_i|_{\epsilon=0}$ if $\epsilon <0$.
This implies that for $\epsilon > 0$, the agent slows down whereas for $\epsilon < 0$, it moves faster.
In other words, the nonlinear term $f(\mathbf{v}_j+\mathbf{v}_i)$ ensures that the
agent's speed remains close to that of the specified mean value.

\clearpage

\section*{Snapshots of flocking patterns observed over range of parameter values}

\begin{figure}[!ht]
\centering
\includegraphics{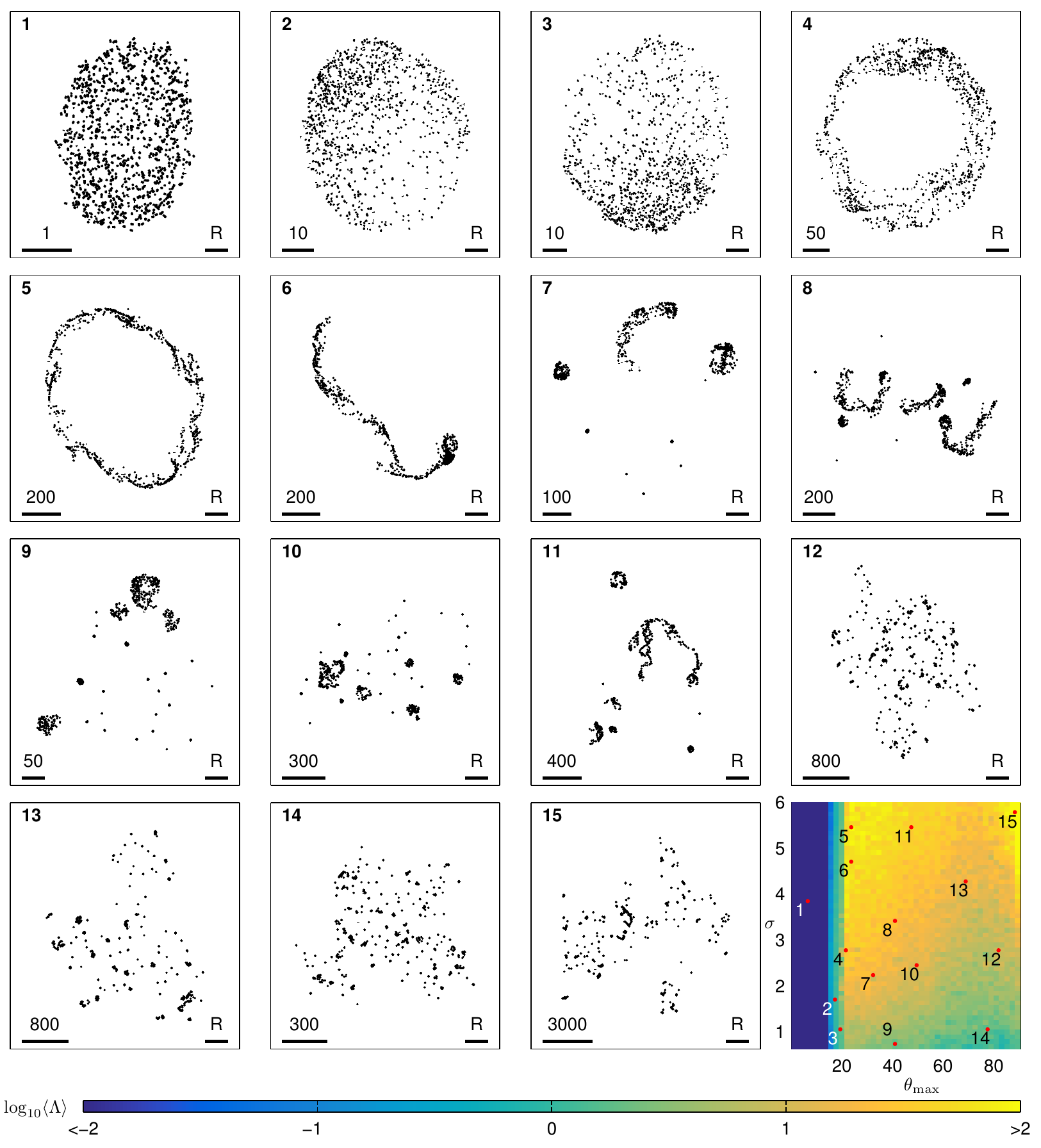}
\caption{
Snapshots of flocking patterns exhibited by the model for a system of $N = 10^{3}$ agents, obtained
for an interaction strength $\alpha=0.01$, over a range of values of the mean interaction length
$\sigma$ and maximum bearing angle $\theta_{\max}$. The corresponding parameter space diagram from
the main text is displayed in the bottom right panel.
In this panel, we display (in log-scale) the dependence of the average angular momentum of the flock
on $\sigma$ and $\theta_{\max}$.
Each of the other $15$ panels display flocking patterns observed for
parameter values denoted by the corresponding numbered red marker on the parameter space diagram.
The numbered solid bars in the lower left corner of these panels provides a measure of spatial
distance in each case. The solid bar in the lower right corner of each panel indicates the extent of
the corresponding resolution length $R$, which we use for our cluster-finding algorithm.
}
\label{snap1}
\end{figure}

\begin{figure}[!ht]
\centering
\includegraphics{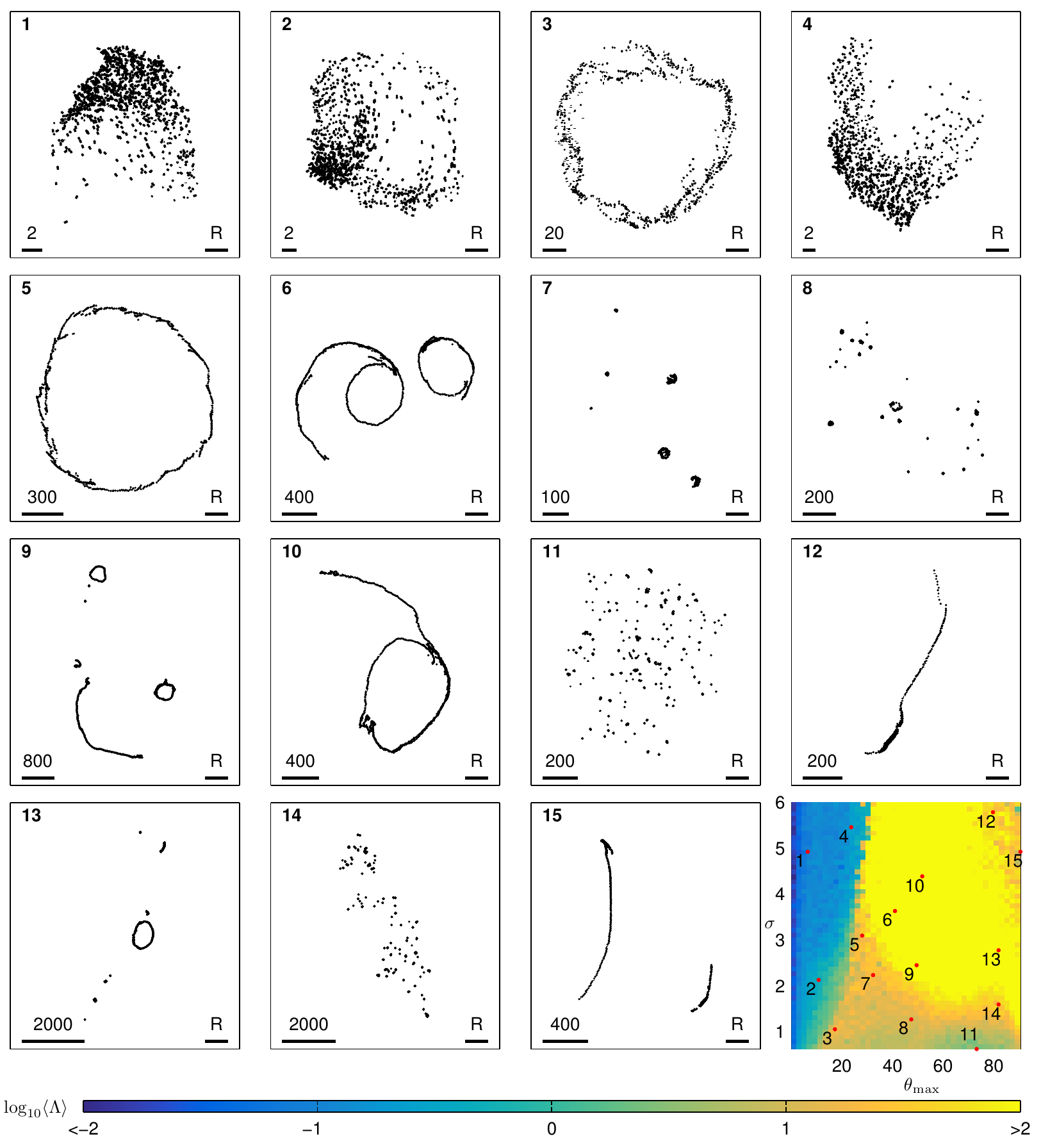}
\caption{
Snapshots of flocking patterns exhibited by the model for a system of $N = 10^{3}$ agents, obtained
for an interaction strength $\alpha=0.05$, over a range of values of the mean interaction length
$\sigma$ and maximum bearing angle $\theta_{\max}$. The corresponding parameter space diagram from
the main text is displayed in the bottom right panel.
In this panel, we display (in log-scale) the dependence of the average angular momentum of the flock
on $\sigma$ and $\theta_{\max}$.
Each of the other $15$ panels display flocking patterns observed for
parameter values denoted by the corresponding numbered red marker on the parameter space diagram.
The numbered solid bars in the lower left corner of these panels provides a measure of spatial
distance in each case. The solid bar in the lower right corner of each panel indicates the extent
of the corresponding resolution length $R$, which we use for our cluster-finding algorithm.
}
\label{snap2}
\end{figure}

\begin{figure}[!ht]
\centering
\includegraphics{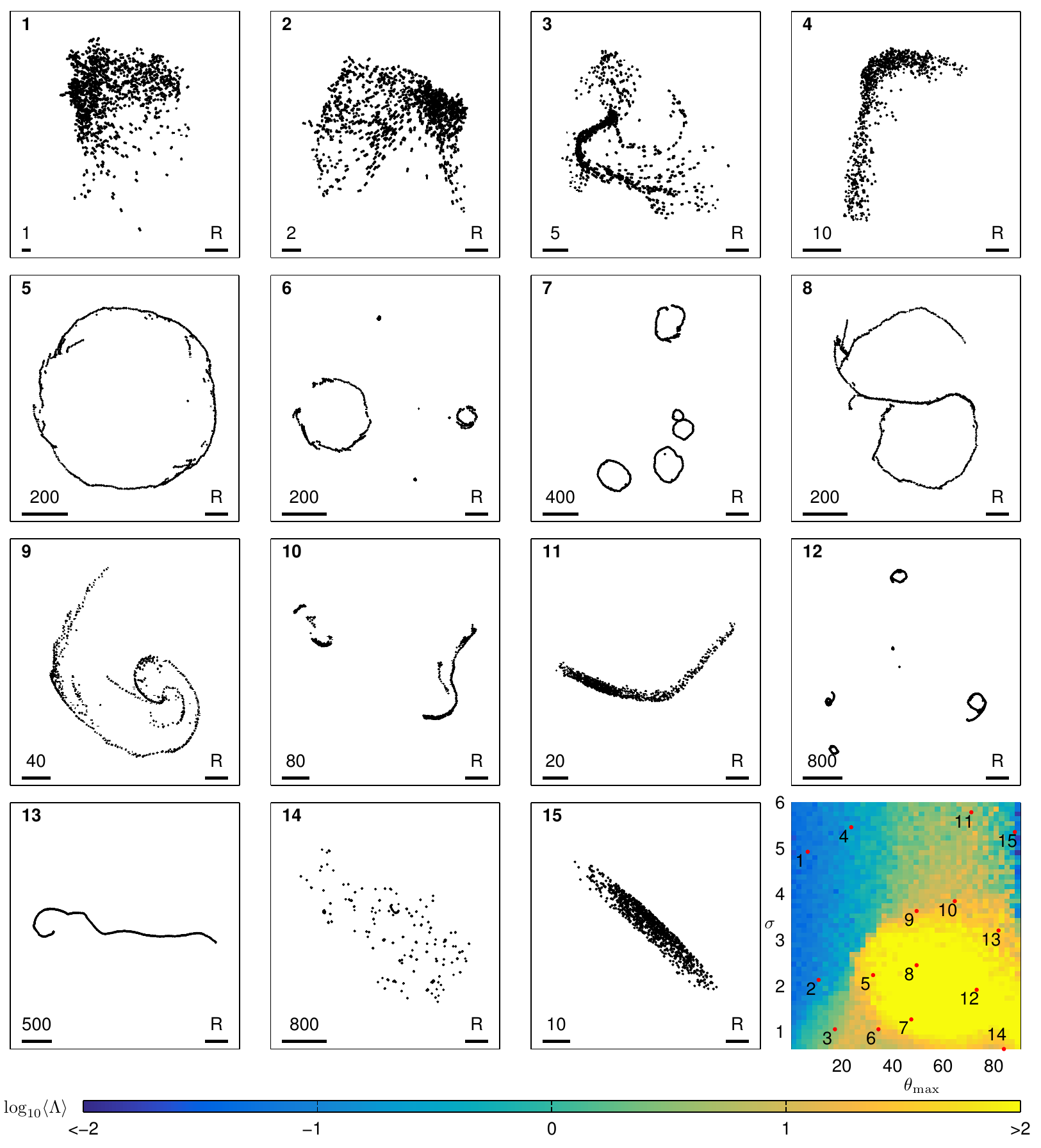}
\caption{
Snapshots of flocking patterns exhibited by the model for a system of $N = 10^{3}$ agents, obtained
for an interaction strength $\alpha=0.1$, over a range of values of the mean interaction length
$\sigma$ and maximum bearing angle $\theta_{\max}$. The corresponding parameter space diagram from
the main text is displayed in the bottom right panel.
In this panel, we display (in log-scale) the dependence of the average angular momentum of the flock
on $\sigma$ and $\theta_{\max}$.
Each of the other $15$ panels display flocking patterns observed for
parameter values denoted by the corresponding numbered red marker on the parameter space diagram.
The numbered solid bars in the lower left corner of these panels provides a measure of spatial
distance in each case. The solid bar in the lower right corner of each panel indicates the extent
of the corresponding resolution length $R$, which we use for our cluster-finding algorithm.
}
\label{snap3}
\end{figure}

\begin{figure}[!ht]
\centering
\includegraphics{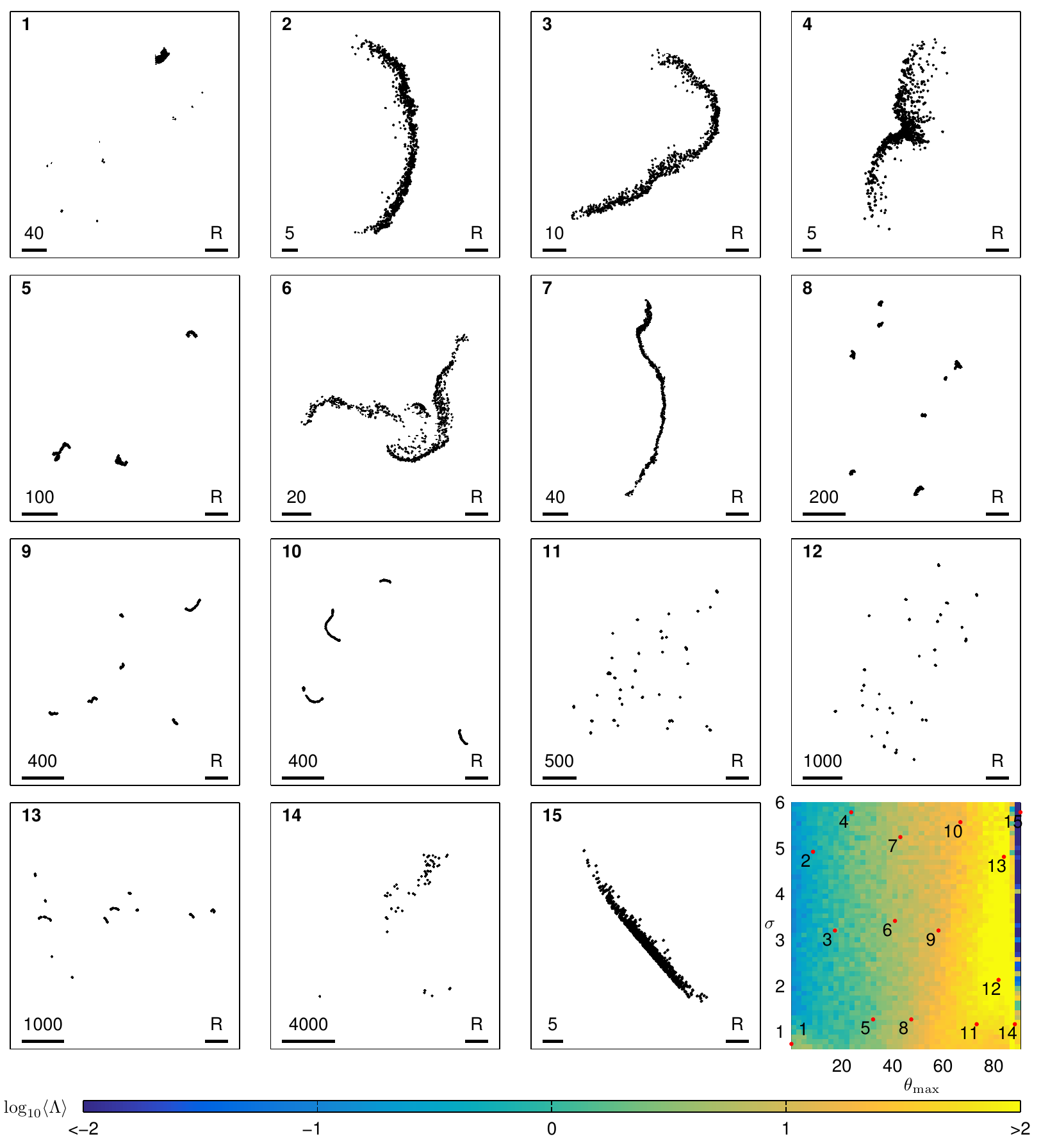}
\caption{
Snapshots of flocking patterns exhibited by the model for a system of $N = 10^{3}$ agents, obtained
for an interaction strength $\alpha=0.5$, over a range of values of the mean interaction length
$\sigma$ and maximum bearing angle $\theta_{\max}$. The corresponding parameter space diagram from
the main text is displayed in the bottom right panel.
In this panel, we display (in log-scale) the dependence of the average angular momentum of the flock
on $\sigma$ and $\theta_{\max}$.
Each of the other $15$ panels display flocking patterns observed for
parameter values denoted by the corresponding numbered red marker on the parameter space diagram.
The numbered solid bars in the lower left corner of these panels provides a measure of spatial
distance in each case. The solid bar in the lower right corner of each panel indicates the extent
of the corresponding resolution length $R$, which we use for our cluster-finding algorithm. Note
that the pattern in panel $1$ is classified as a single cluster because over $90\%$ of the agents
belong to that cluster (see algorithm for details). Moreover while the snapshots of patterns in
panels $2-4$ may appear reminiscent of the wriggling pattern, their dynamics are in fact qualitatively
similar to the meandering pattern.
}
\label{snap4}
\end{figure}

Flocking patterns observed for $\alpha=0.01,0.05,0.1,0.5$,
and over a range of $\theta_{\max}$ and $\sigma$, are displayed in Figs.~\ref{snap1}--\ref{snap4}.

\clearpage

\clearpage

\clearpage

\clearpage

\section*{Description of the movies}

The captions for the four movies are displayed below:

\begin{itemize}
\item
\begin{verbatim}
Movie_S1.mp4
\end{verbatim}
Evolution of a system of $N=10^{3}$ agents moving in a wriggling pattern for the case
$\sigma=5$, $\theta_{\max}=40$ and $\alpha=0.8$. The system is simulated over
$2\times10^{4}$ time steps, starting from an initial condition where agents are
distributed randomly over a small portion of the computational domain. Each frame of
the simulation is separated by $50$ time steps.

\item
\begin{verbatim}
Movie_S2.mp4
\end{verbatim}
Evolution of a system of $N=10^{3}$ agents moving in a closed trail for the case
$\sigma=3$, $\theta_{\max}=50$ and $\alpha=0.1$. The system is simulated over
$2\times10^{4}$ time steps, starting from an initial condition where agents are
distributed randomly over a small portion of the computational domain. Each frame of
the simulation is separated by $50$ time steps.

\item
\begin{verbatim}
Movie_S3.mp4
\end{verbatim}
Evolution of a system of $N=10^{3}$ agents moving in a milling pattern for the case
$\sigma=1$, $\theta_{\max}=20$ and $\alpha=0.025$. The system is simulated over
$2\times10^{4}$ time steps, starting from an initial condition where agents are
distributed randomly over a small portion of the computational domain. Each frame of
the simulation is separated by $50$ time steps.

\item
\begin{verbatim}
Movie_S4.mp4
\end{verbatim}
Evolution of a system of $N=10^{3}$ agents moving in a flock with a meandering
center of mass for the case $\sigma=3$, $\theta_{\max}=15$ and $\alpha=0.02$. The
system is simulated over $2\times10^{4}$ time steps, starting from an initial
condition where agents are distributed randomly over a small portion of the
computational domain. Each frame of the simulation is separated by $50$ time steps.

\end{itemize}


\end{document}